\documentstyle[manuscript,epsf,aps,floats]{revtex}

\newcommand{\bracket}[1]{\langle #1 \rangle}    
\newcommand{\prt}[1]{\left( #1 \right)}         

\def\unafigura#1#2#3#4{  
\begin{figure} 
\centerline{\mbox{\epsfxsize=#4 \epsfbox{#1} }}
\caption{#2}
\label{#3}
\end{figure}
}

\def\duesfiguresvert#1#2#3#4#5#6{  
\begin{figure}[p]
\begin{center}
  \mbox{\epsfxsize=10cm \epsfbox{#1} }
  \caption{#3}\label{#5}
  \vspace{2cm}
  \mbox{\epsfxsize=10cm \epsfbox{#2} }
  \caption{#4}\label{#6}
\end{center}
\end{figure}
}

\newcommand{\figloopshisto}{
\unafigura{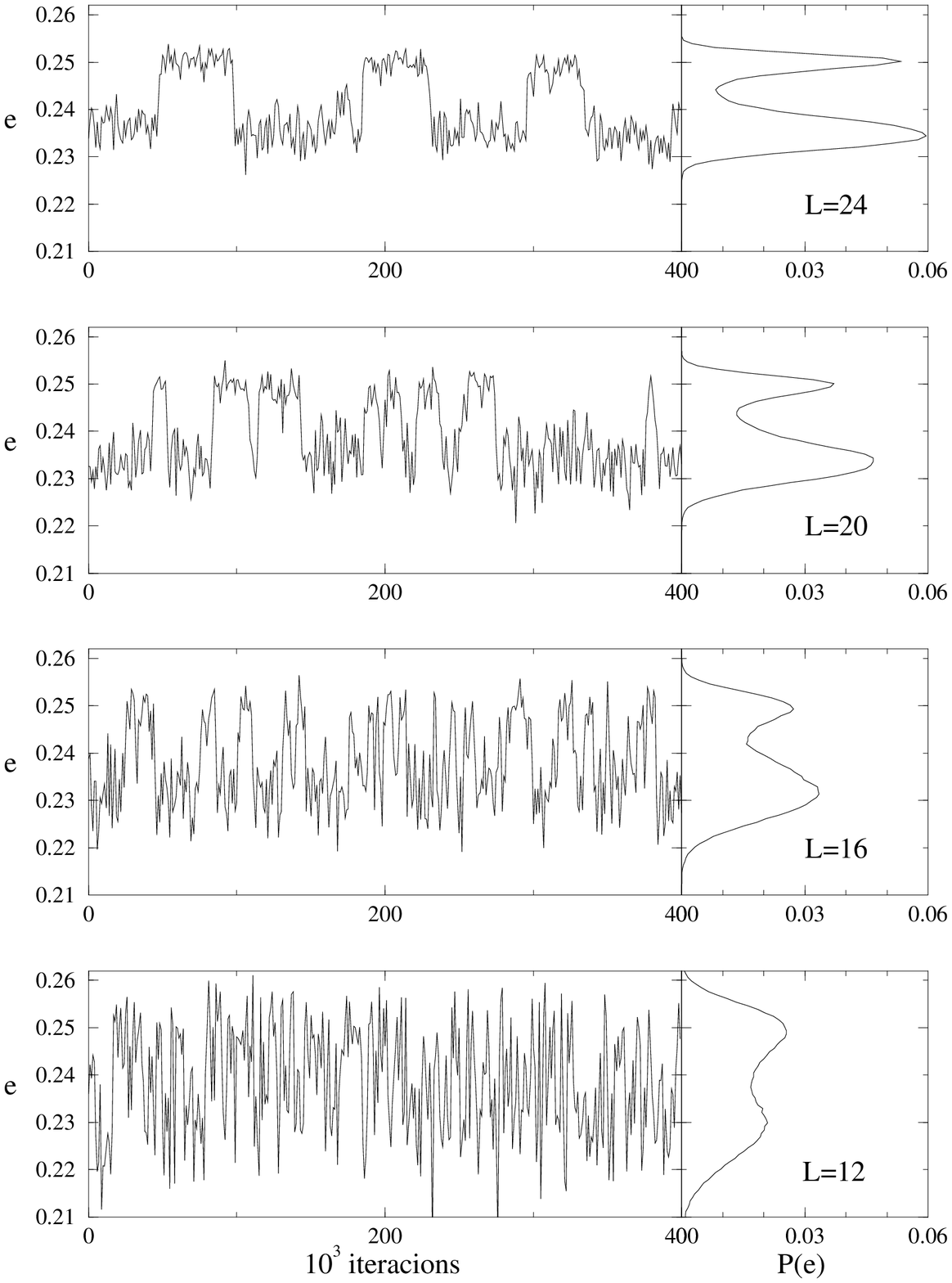}%
{Some Monte Carlo runs with their corresponding histograms for
  $L=12,16,20,24$. All the axes are equally scaled in order to compare them.
  Notice that for large $L$ the separation between phases grows, that is, the
  two peaks structure is more clear and in the time-series, the chance for
  tunneling decreases.}
{fig:histograms}{12cm} }

\newcommand{\figlooopsFitNuGap}{
\duesfiguresvert{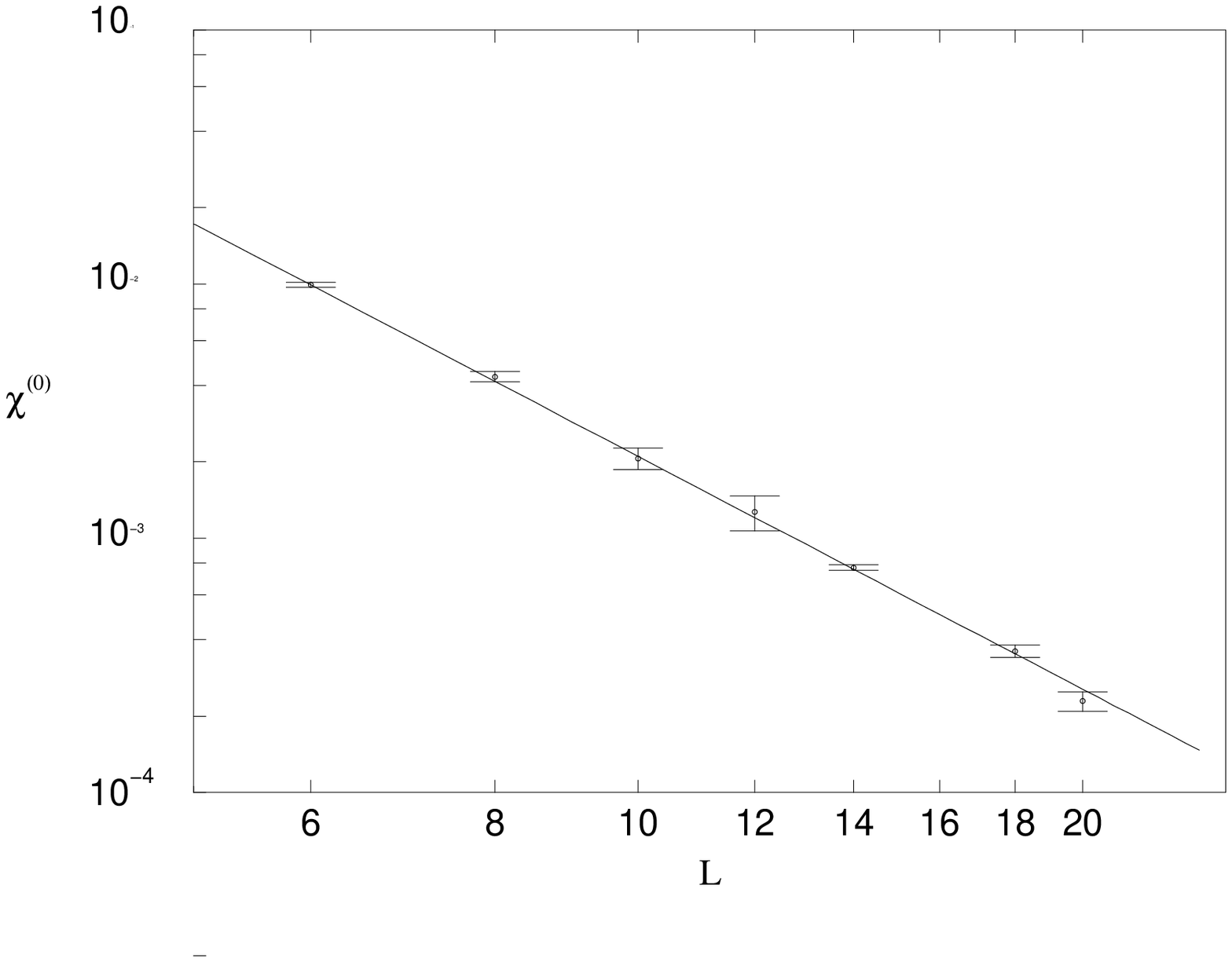}{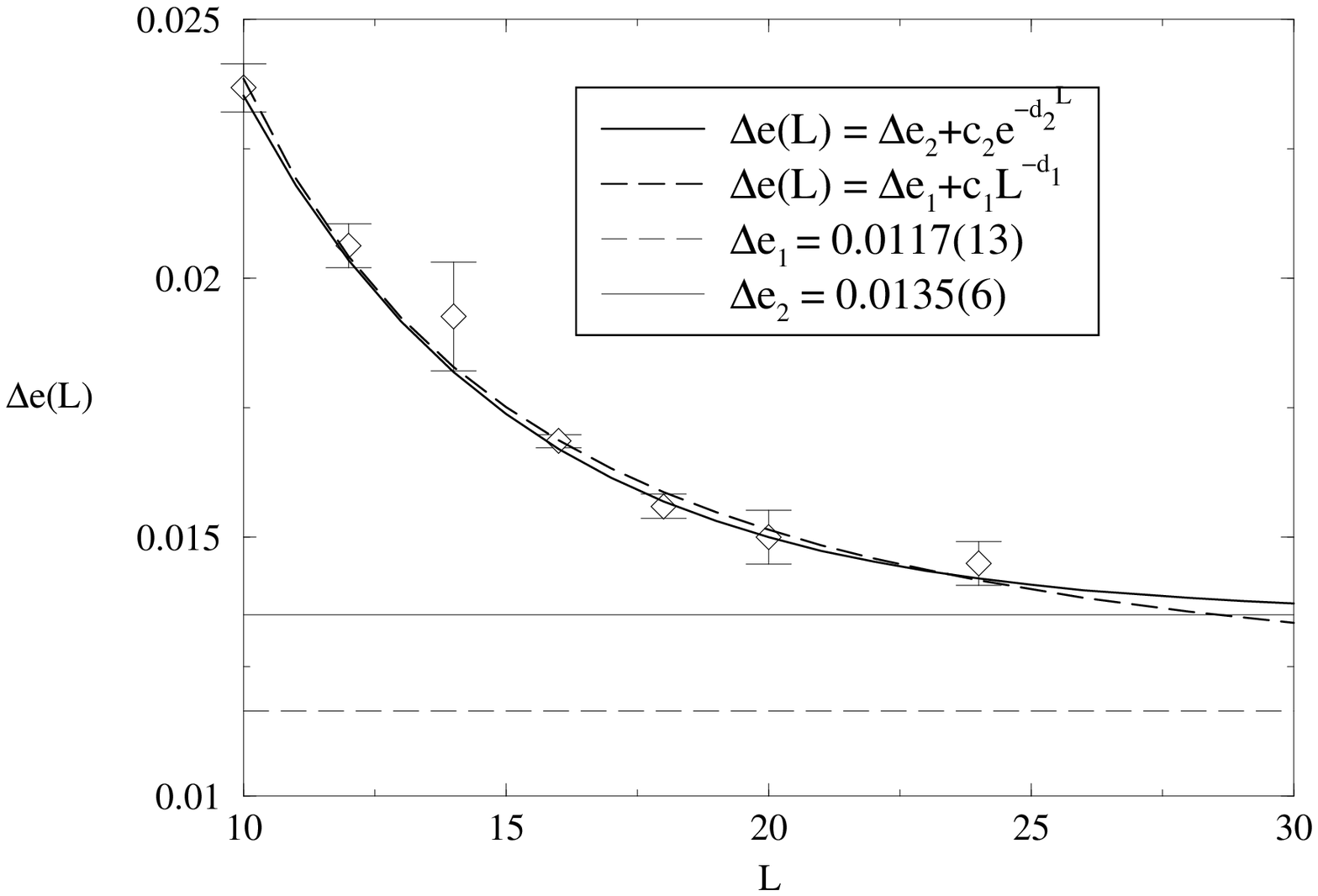}
{FSS of $\chi^{(0)}$, the imaginary part of the zeroes in the partition
  function. This is a log-log plot, so the fit must be a line (the solid line)}
{FSS of the latent heat. Data fit so well an exponential
  law~(\ref{eq:loops:gap_exp}) and a potential one~(\ref{eq:loops:gap_power}).
}{fig:loops:FitNu}{fig:loops:gap}
}

\newcommand{\figlooopsCvMax}{
\unafigura{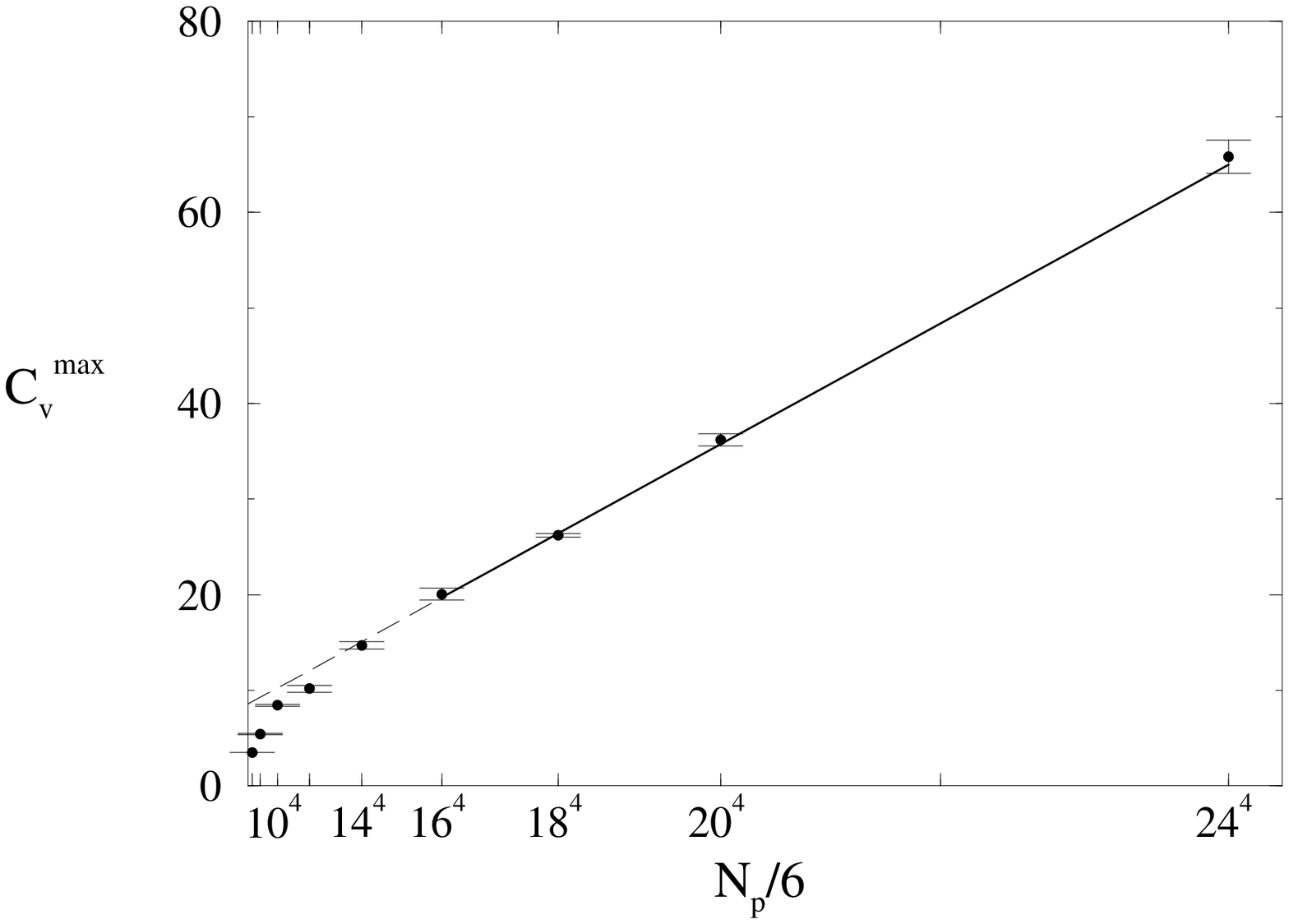}
{FSS of the specific heat as a function of the volume. 
The points with error bars
  are the data and the solid line is the fit of
  eq.~(\ref{eq:cmax1}) using the three last points. Notice that the smaller
  lattices are away from this fit.}
{fig:loops:CvMax}{10cm}
}

\begin{document}

\draft 
\tightenlines 
\preprint{\vbox{
\hbox{UAB--FT/445}
\hbox{hep--lat/9807045}
}}


\title{
Finite Size Analysis of the $U(1)$
Phase Transition using the World-sheet
Formulation
}

\author{ M.~Baig and J.Clua}
\address{Grup de F{\'\i}sica Te{\`o}rica-IFAE, Universitat Aut{\`o}noma de Barcelona \\
         08193 Bellaterra (Barcelona) Spain.}
\author{H.~Fort}
\address{Instituto de F{\'\i}sica, Facultad de Ciencias, Universidad de la
        Rep{\'u}blica.\\
        Tristan Narvaja 1674, 11200 Montevideo, Uruguay.}

\vspace{2cm}
\date{ July 31, 1998 }

\maketitle 

\begin{abstract}
  We present a high statistics analysis of the pure gauge compact U(1) lattice
  theory using the the $world-sheet$ or Lagrangian $loop$ representation.  
  We have applied a
  simulation method that deals directly with (gauge invariant) integer 
  variables on plaquettes. As a
  result we get a significant amelioration of the simulation that allows to
  work with large lattices avoiding the metaestability problems that appear
  using the standard Wilson formulation.
\end{abstract}



\section{Introduction}

It is well known that pure $U(1)$ lattice gauge theory exhibits a phase
transition separating a region where photons remain confined (strong coupling
regime) and a Coulombian region (weak coupling).  Nevertheless, the order of
such transition is rather controversial~\cite{Campos:1998jp,Campos:1997br}. %
Simulations on hypercubical lattices using the Wilson form of the action shown
the existence of a latent heat that seems to persist in the infinite volume
limit~\cite{Klaus:1997be,Roiesnel:1997am}.%
If this transition is actually of first order, hence the gauge U(1)
lattice theory would be lacking of a continuum limit.

The seriousness of such a situation has motivated the appearance of two lines
of study. The first one is intended to check if the unexpected first order
nature of this phase transition is an ``accident'' of the simulation,
originated by the imposition of an hypercubical lattice~\cite{Jersak:1996mj},
the periodic boundary conditions~\cite{Baig:1994ia}, etc. The second line of
work goes directly to the the physical origin of the phase transition, the
monopole condensation~\cite{Banks:1977cc}, and to the analysis of all related
effects, as the presence of monopole percolation~\cite{Baig:1994ib}, that can
shadow the behavior of the phase transition.

Although the formulation of the compact pure gauge U(1) theory on a lattice is
very simple, i.e. there are only phases associated to the links, obtaining
reliable numerical results from simulations is a rather difficult task. The
earlier numerical simulations of such a theory indicated actually a continuous
phase transition~\cite{Lautrup:1980xr,Bhanot:1981zg}. The evidences which
favor a first order transition appeared only when larger lattices and higher
statistics were accessible~\cite{Jersak:1983yz,Azcoiti:1990ue,Bhanot:1992nf}.%
However, numerical simulations on larger lattices are very difficult. This is
originated mainly by the lack of tunneling between phases that become
incredibly stable and difficult the statistical analysis.

The ordinary Wilson formulation of QED has the links as the basic variables
representing the gauge field and therefore involves a gauge redundancy.  A
different implementation of QED on the lattice which avoid the gauge
redundancy can be attained using a description directly in terms of the
world-sheets of gauge invariant quantities or loop excitations. This is the
Lagrangian counterpart of the former introduced Hamiltonian loop
representation~\cite{Gambini:1980wm,Gambini:1986ew}.  Although the
non-canonical nature of the loop algebra had made elusive this Lagrangian
formulation, in a previous set of papers~\cite{Aroca:1994yb,Aroca:1996cs} a
procedure to cast loops into the so-called {\em world-sheet formulation}
was introduced.

In this paper we present a large statistics analysis of the pure gauge
U(1) lattice theory using the world-sheet formulation. The lattice action used
is equivalent to the Villain form. In this formulation basic variables
are integer variables on the plaquettes.  
As we will comment, the thermodynamical
behavior over the phase transition is different to the one obtained 
from a standard simulation.  The persistent metaestability problems that appear
in the simulations using the traditional formulation are absent in the case
of the world-sheet formulation . This therefore makes this formulation 
a useful computational tool. 

The plan of the paper is as follows. In Sec. II we summarize the basics of the
world-sheet approach. In Sec. III we present the simulation method and the
results using the world-sheet action corresponding to the Villain U(1) theory.
Sec. IV is devoted to the analysis of the critical exponents of the transition.
Finally, sec. V contains the summary of the work and the main conclusions.

\section{The Lagrangian Loop Approach or the World-Sheet Formulation
\label{sec:loops:lagrangia} }

QCD is generally defined in terms of local fields, quarks and gluons, but the
physical excitations are extended composites: mesons and baryons.  There is an
alternative quantum formulation of Yang-Mills theories directly in terms of
these extended excitations namely, the {\em loop
  representation}~\cite{Gambini:1980wm,Gambini:1986ew}. %
The basis of the loop description of gauge theories can be traced to the idea
of describing gauge theories explicitly in terms of Wilson loops or
holonomies~\cite{Polyakov:1979gp,Polyakov:1980ca,Makeenko:1979pb} since
Yang~\cite{Yang:1974kj} noticed their important role for a complete
description of gauge theories.  Loops replace the information furnished by
the vector potential.  A description in terms of loops or strings, besides the
general advantage of only involving the gauge invariant physical excitations,
is appealing because all the gauge invariant operators have a simple
geometrical meaning when realized in the loop space.

The loop based approach of ref.~\cite{Gambini:1980wm,Gambini:1986ew} describes
the quantum electrodynamics in terms of the gauge invariant holonomy (Wilson
loop)
\begin{equation}
\hat{W} (\gamma ) = \exp [i e \oint_{\gamma} A_a (y) dy^a] , 
\label{eq:Wloop}
\end{equation}
and the conjugate electric field $\hat{E}^a (x)$ .  They obey the commutation
relations
\begin{equation}
[\hat{E}^a(x), \hat{W} (\gamma)] = e \int_\gamma \delta(x - y) dy^a \hat{W}
(\gamma) .
\label{eq:alg}
\end{equation}

These operators act on a state space of Abelian loops $\psi(\gamma)$ that may
be expressed in terms of the transform

\begin{equation}
\psi (\gamma) = \int d_\mu [A] <\gamma \mid A> <A \mid \psi>
= \int d_\mu [A] \psi [A] \exp [- i e \oint_{\gamma} A_a dy^a] .
\label{eq:trans}
\end{equation}

This loop representation has many appealing features: first, it allows to do
away with the first class constraints of gauge theories.  That is, the Gauss
law is automatically satisfied. Second, the formalism only involves gauge
invariant objects. Third, all the gauge invariant operators have a transparent
geometrical meaning when they are realized in the loop space.

When this loop representation is implemented in the lattice it offers a gauge
invariant description of physical states in terms of kets $\mid C >=\hat{W}(C)
\mid 0>$, where $C$ labels a closed path in the {\em spatial} lattice.
Eq.(\ref{eq:alg}) becomes
\begin{equation}
[\hat{E}_l,\hat{W}(C)] =  N_l(C)\hat{W}(C),
\label{eq:alg1}
\end{equation}
where $l$ denotes the links of the lattice, $\hat{E}(l)$ the lattice electric
field operator, $\hat{W}(C)=\prod_{l\in C}\hat{U}(l)$ and $N_l(C)$ is the
number of times that the link $l$ appears in the closed path $C$.

In this loop representation, the Wilson loop acts as the loop creation
operator:
\begin{equation}
\hat{W}(C')\mid C> = \mid C'\cdot C>.
\label{eq:Wloop1}
\end{equation}
The physical meaning of a loop may be deduced  from
(\ref{eq:alg1}) and (\ref{eq:Wloop1}), in fact
\begin{equation}
\hat{E}_l \mid C> = N_l(C) \mid C>,
\label{eq:E}
\end{equation}
which implies that $\mid C>$ is an eigenstate of the electric field. The
corresponding eigenvalue is different from zero if the link $l$ belongs to
$C$. Thus $C$ represents a line of electric flux.

In order to cast the loop representation in Lagrangian form it is convenient
to use the language of differential forms on the lattice of
ref.~\cite{Guth:1980gz}.  Besides the great simplifications to which this
formalism leads its advantages consists in the general character of the
expressions obtained.  That is, most of the transformations are independent of
the space-time dimension or of the rank of the fields. So let us summarize the
main concepts and some useful results of the differential forms formalism
on the lattice.

A k-form is a function defined on the k-cells of the lattice (k=0 sites, k=1
links, k=2 plaquettes, etc.) over an Abelian group which shall be {\bf R},
{\bf Z}, or U(1)={reals module 2$\pi$}.  Integer forms can be considered
geometrical objects on the lattice.  For instance, a 1-form represents a path
and the integer value on a link is the number of times that the path traverses
this link.  $\nabla$ is the co-border operator which maps k-forms onto
(k+1)-forms. It is the gradient operator when acting on scalar functions
(0-forms) and it is the rotational on vector functions (1-forms). We shall
consider the scalar product of p-forms defined $<\alpha \mid \beta> =
\sum_{c_k}\alpha (c)\beta (c)$ where the sum runs over the k-cells of the
lattice. Under this product the $\nabla$ operator is adjoint to the border
operator $\partial$ which maps k-forms onto (k-1)-forms and which corresponds
to minus times the usual divergence operator. That is,

\begin{eqnarray}
<\alpha \mid  \nabla \beta> = <\partial \alpha \mid \beta>,\\
<\nabla \alpha \mid \beta> = <\alpha \mid \partial \beta>.
\label{eq:inter}
\end{eqnarray}

The co-border $\nabla$ and border $\partial$ operators verify
\begin{equation}
{\nabla}^2 = 0,\;  \;  {\partial}^2 = 0.
\label{eq:0sq}
\end{equation}

The Laplace-Beltrami operator operator is defined by  
\begin{equation}
\Box =\nabla \partial +\partial \nabla .
\label{eq:Box}
\end{equation}
It is a symmetric linear operator which commutes with $\nabla$ and $\partial$,
and differs only by a minus sign from the current Laplacian $ \Delta_\mu
\Delta_\mu$.  From Eq.(\ref{eq:Box}) is easy to show the Hodge-identity:

\begin{equation}
1=\partial {\Box}^{-1}\nabla  + \nabla {\Box}^{-1}\partial .
\label{eq:Hodge}
\end{equation}





After this parenthesis on differential forms on the lattice let us consider
the generating functional for the Wilson lattice U(1) action:
\begin{equation}
Z_W = \int_{-\pi}^{\pi} (d\theta_l )\exp(-\frac{\beta}{2}\sum_p \cos 
{\theta}_p),
\end{equation}
where the subscripts $l$ and $p$ stand for the lattice links and plaquettes
respectively.

Fourier expanding the $\exp [ \cos \theta ]$ we get
\begin{equation}
Z_W = \int_{-\pi}^{\pi} (d\theta_l )  \prod_p \sum_{n_p} I_{n_p}(\beta ) 
e^{in_p {\theta}_p},
\end{equation}
which can be written, using the differential forms language as
\begin{equation}
Z_W = \sum_{ \left\{ n_p \right\} }
\int_{-\pi}^{\pi} (d\theta_l ) \exp (\sum_p \ln I_{n_p}(\beta ) ) 
e^{i<n , \nabla \theta_l >}.
\end{equation}
In the above expression, $\theta_l$ is a real periodic 1-form, that is, a real
number $\theta \in [-\pi, \pi ]$ defined on each link of the lattice; $\nabla$
is the co-border operator; $n_p$ are integer 2-forms, defined at the lattice
plaquettes.  By eq. (\ref{eq:inter}) and integrating over $\theta_l$ we obtain
a $\delta (\partial n_p)$.  Then, the partition function can be written as
\begin{equation}
Z_W \propto \sum_{ \left\{ n_p; \partial n_p = 0 \right\} }
\exp (\sum_p \ln I_{n_p}(\beta ) ), 
\label{eq:Wil}
\end{equation}
the constraint $\partial n_p = 0$ means that the sum is restricted to $closed$
2-forms. Thus, the sum runs over collections of plaquettes constituting closed
surfaces.

An alternative and more easy to handle lattice action than the Wilson action
is the Villain action. The partition function for this action is given by
\begin{equation}
Z_V = \int (d\theta ) \sum_{ s }\exp
(-\frac{{\beta}_V}{2}\mid\mid \nabla \theta -2\pi s\mid\mid^2),
\label{eq:Villain}
\end{equation}
where $\mid \mid \ldots \mid \mid^2 = <\ldots , \ldots>$.  If we use the
Poisson summation formula
$$\sum_s f(s) = \sum_n \int_{-\infty}^{\infty} d\phi 
f(\phi) e^{2\pi i\phi n}$$
and we integrate the continuum $\phi$ variables we get
\begin{equation}
Z_V = (2\pi {\beta}_V)^{-N_p/2} \int (d\theta ) \sum_{n }\exp
(-\frac{1}{2{\beta}_V}<n,n>+i<n,\nabla \theta >),
\label{}
\end{equation}
where $N_p$ in the number of plaquettes of the lattice. Again, we can use the
equality: $<n,\nabla \theta>=<\partial n,\theta >$ and integrating over
$\theta$ we obtain a $\delta (\partial n)$. Hence we get:
\begin{equation}
Z_V = (2\pi {\beta}_V)^{-N_p/2}  \sum_{\left\{ n; \partial n = 0 \right\}}
\exp(-\frac{1}{2{\beta}_V}<n,n>),
\label{eq:Vill}
\end{equation}
where $n$ are integer 2-forms. Eq. (\ref{eq:Vill}) is obtained from
Eq.(\ref{eq:Wil}) in the $\beta \rightarrow \infty$ limit.

If we consider the intersection of one of such surfaces with a $t=constant$
plane we get a loop $C_t$. It is easy to show that the creation operator for
this loop is just the creation operator for the loop representation i.e.  the
Wilson loop operator. Repeating the steps from Eq.(\ref{eq:Villain}) to
Eq.(\ref{eq:Vill}) we get for $<\hat{W}(C_t)>$
\begin{equation}
<W(C_t)> = \frac{1}{Z}(2\pi {\beta}_V)^{-N_p/2}  
\sum_{
                \begin{array} {c} n\\
                              \left(  \partial n = C_t \right)
                \end{array}
                 }\exp
(-\frac{1}{2{\beta}_V}<n,n>).
\label{eq:W}
\end{equation}
This is a sum over all world-sheets spanned on the loop $C_t$.  In other
words, we have arrived to an expression of the partition function of compact
electrodynamics in terms of the world-sheets of pure electric loop
excitations.  Notice that the loop action corresponding to the Villain action
is proportional to the $quadratic$ $area$ $A_2$:
\begin{equation}
S_L = -\frac{1}{2{\beta}_V} A_2 = 
-\frac{1}{2{\beta}_V}\sum_{p \in {\cal S}} n_p^2 = -\frac{1}{2{\beta}_V}<n,n>,
\label{eq:A2}
\end{equation}
i.e. the sum of the squares of the mul\-ti\-pli\-ci\-ties $s_p$ of
pla\-que\-ttes which constitute the loop's world sheet ${ \cal S}$. It is
interesting to note the similarity of this action with the continuous Nambu
action or its lattice version, the Weingarten action~\cite{Weingarten:1980gn}, which is
proportional to the area swept out by the bosonic string.

\section{Numerical analysis\label{sec:loops:detalls}}

We simulated the loop lattice action of eq. (\ref{eq:A2}).  In
ref.~\cite{Aroca:1994yb} we shown that this action is equivalent to the
Integer Gaussian Model (IGM) that, in turn, it can be also reached as a limit
of the non compact scalar QED in the limit of infinite Higgs coupling. In that
paper we presented a direct simulation of the IGM which from the analysis of
the energy histograms showed clearly the first order nature of the transition.
Nevertheless, we did not computed the critical exponents of the phase
transition.

Here we have performed a more quantitative study and computed the exponents of
the phase transition. We applied the simulation method introduced in
ref.\cite{Aroca:1996cs}. This method consists in generating collections of integer
variables attached to plaquettes which form surfaces with the appropriate
constraints for each model (null frontier in the present pure gauge case). 
The variables are updated by means of a standard Metropolis Monte Carlo algorithm.
These larger simulations produced  a phase transition for the coupling constant
$\beta = 1/(2\beta_V) = 0.778$ i.e. $\beta_V = 1/(2* 0.778) = 0.6426$ which is
slightly higher than the previous value of $\beta_V \simeq 0.64$ 
\cite{Aroca:1996cs}.

The physical magnitudes measured under the simulation are the internal energy
and the specific heat both normalized to the number of plaquettes $N_p = 6
L^4$:
\begin{eqnarray}
  e   & = & N_p^{-1} \bracket{E}                  \\
  C_v & = & \beta^2 N_p^{-1} 
            \prt{\bracket{E^2} - \bracket{E}^2 }
\end{eqnarray}
on $\beta \equiv {1}/{2\beta_V}$.

The details of the run are the following:
\begin{itemize}
\item Hypercubical lattice (that is, periodic boundary conditions)
\item Sizes ranging from $L=8$ to $L=24$.
\item Good statistics,  more than 100 times the self-correlations. It
  means millions of iterations for $L=24$.
\item In order to locate the transition and the specific heat peaks, we didn't
  do thermal cycles. Instead we used iteratively Ferrenberg-Swedsen reweighting
  until a two peaks histogram were found.
\end{itemize}

We made a FSS analysis for the following quantities:
\begin{itemize}
\item[$\diamondsuit$] The specific heat maxima
  $C_{max}(L)$~(\S\ref{subsec:loops:critic:Cv}).
\item[$\diamondsuit$] The critical coupling position
  $\beta_c(L)$~(\S\ref{subsec:loops:critic:BetaCrit}).
\item[$\diamondsuit$] The partition function zeroes in the complex
  plane.~(\S\ref{subsec:loops:critic:LeeYang}).
\item[$\diamondsuit$] The latent heat, that is the distance between the two
  peaks~(\S\ref{subsec:loops:critic:DeltaE}).
\end{itemize}

The calculations have been performed in Pentium computers running Linux
(including the RTNN machine of 32 Pentium Pro processors at 200MHz).
As it is expected, the elapsed time in the runs grows with the volume of the
lattice obeying the following relation:
\begin{equation}
  t = 42 \prt{\frac{L}{10}}^4 \mbox{hours/Mit},
  \label{eq:loops:CPUTime}
\end{equation}
where $t$ is the elapsed time for a Pentium Pro 200MHz running a
million of iterations.

We have been very careful with the statistical analysis of such
a transition. We have measured the autocorrelations, $\tau_e$, of the
time-series for each lattice size in order to determine both, the number of
thermalization iterations and the number of measures to obtain a significative
number of independent measures. It is accepted \cite{Sokal:1989ea} that about
$20\tau_e$ iterations are enough in order to have the series thermalized and
about $100\tau_e$ to have a sufficiently large time series.
Table~{\ref{tab:correlations}} collects this data.
\begin{table}[htbp]
  \begin{center}
    \begin{tabular}[c]{rrrrr}
      \multicolumn{1}{c}{$L$} &
      \multicolumn{1}{c}{$\beta_{MC}$} &
      \multicolumn{1}{c}{$\tau$} &
      \multicolumn{1}{c}{$n_{\rm{th}}/\tau$} &
      \multicolumn{1}{c}{$n_{\rm{m}}/\tau$} \\
      \hline
       6 & 0.77991 &   90 & 1100 & 110000 \\
       8 & 0.77821 &  300 &  330 &    660 \\
      10 & 0.77821 &  500 &  100 &    200 \\
      12 & 0.77700 &  810 &   70 &    140 \\
      14 & 0.77700 & 1500 &   60 &    100 \\ 
      16 & 0.77688 & 2100 &   50 &    225 \\
      18 & 0.77688 & 3300 &   40 &    130 \\ 
      20 & 0.77688 & 5000 &   30 &    110 \\
      24 & 0.77680 &13100 &   20 &    100 \\ 
    \end{tabular}
    \caption{Correlations}
    \label{tab:correlations}
  \end{center}
\end{table}
Using this data, one obtains that the self-correlation time grows
exponentially (critical slowing down) according to
\begin{equation}
  \tau = \tau_0 e^{2\sigma L} \qquad \tau_0=49(4)
                               \quad \sigma=0.116(2)
                               \quad {\cal Q}=0.95,
  \label{eq:loops:CritSlowingDown}
\end{equation}
where we supposed an error of about a 5\% measuring $\tau$.\footnote{It is
  difficult to give a realistic estimation for the error in
  $\tau$. Nevertheless, even underestimating the errors ~(1\%) the goodness of
  the fit is not bad at all (${\cal Q}=0.04$). }

The goodness of the fits is done using
\begin{equation}
 {\cal Q} = Q\prt{\frac{d.o.f}{2},\frac{\chi^2}{2}}
\label{eq:errors:Q}
\end{equation}
where $Q(a,x)$ is the regularized gamma function 
\begin{equation}
  Q(a,x) \equiv 1 - P(a,x)
         \equiv \frac{\Gamma(a,x)}{\Gamma(a)} 
         \equiv \frac{1}{\Gamma(a)}\int_x^\infty e^{-t} t^{a-1} dt.
\end{equation}

Combining~(\ref{eq:loops:CPUTime}) and~(\ref{eq:loops:CritSlowingDown}) we
obtain the CPU times needed for an arbitrary simulation.
We spent about three months of calculations for $L=24$, and a simple estimation
reveals that to calculate $L=28$ in our machine we
would need more than a year.

In Fig.~\ref{fig:histograms} one can see energy histograms for different
lattice sizes an the time series showing the presence of tunneling between the
two phases. This undoubtedly shows that the
world-sheet formulation (in terms of integer variables on plaquettes)
is considerable more effective than the standard one (in terms
of gauge potentials on links) to carry out a simulation.

\figloopshisto

\section{Critical exponents\label{sec:loops:critic}}

Table~\ref{tab:loops:Observables} summarizes all the observables measured in
  this work.

\begin{table}[htbp]
  \begin{center}
    \begin{tabular}[c]{rrrrlr@{.}ll}
      \multicolumn{1}{c}{$L$} &
      \multicolumn{1}{c}{$\beta_{MC}$} &
      \multicolumn{1}{c}{$\beta_c$} &
      \multicolumn{1}{c}{$\eta^{(0)}$} &
      \multicolumn{1}{c}{$10^{-3} \cdot \chi^{(0)}$} &
      \multicolumn{2}{c}{$C_v^{max}$} &
      \multicolumn{1}{c}{$\Delta e$}\\
      \hline
       6 & 0.77991 & 0.7806(5) & 0.780215(2) & 9.9(2)   &  3&4750(7) & -   \\
       8 & 0.77821 & 0.7785(4) & 0.778358(3) & 4.3(1.8) &  5&42(6)   & -   \\
      10 & 0.77821 & 0.7780(2) & 0.777978(4) & 2.0(2)   &  8&44(9)   & 0.0237(5) \\
      12 & 0.77700 & 0.7770(2) & 0.776985(1) & 1.2(2)   & 10&2(3)    & 0.0206(4) \\
      14 & 0.77700 & 0.7769(7) & 0.776972(8) & 0.76(2)  & 14&7(4)    & 0.019(1)  \\
      16 & 0.77699 & 0.7769(5) &  -          &  -       & 20&1(5)    & 0.0169(1) \\
      18 & 0.77688 & 0.7768(2) & 0.776843(2) & 0.36(3)  & 24&2(2)    & 0.0155(2) \\
      20 & 0.77688 & 0.7768(8) & -           &  -       & 36&2(7)    & 0.0160(5) \\
      24 & 0.77680 & 0.7767(7) &  -          & 0.23(2)  & 65&8(1.8)  & 0.0145(4) \\ 
    \end{tabular}
    \caption{Observables measured in this work. 1st an 2nd column contain the
      size of the lattice and $\beta_{MC}$ of the run. 3rd column contains the
      value of the coupling, $\beta_c$, of the maximum of $C_v$. In 4th and
      5th columns one can find the real and imaginary part of the coupling
      that vanishes $Z$. 6th column contains $C_v$ and finally 7th column is
      the latent heat,  $\Delta e$.}
    \label{tab:loops:Observables}
  \end{center}
\end{table}

\subsection{Specific heat\label{subsec:loops:critic:Cv}}
In a 1st order phase transition, the specific heat is expected to have a
volume scaling law,  {\em i.e.}
\begin{equation}
C_v^{max}(L)=a_1+ b_1 L^D.
\label{eq:cmax1}
\end{equation}
But a continuous phase transition must obey
\begin{equation}
C_v^{max}(L)=a_2+b_2L^{\alpha/\nu}.
\label{eq:cmax2}
\end{equation}
Both expressions are equal if $\alpha=1$ and $\nu=1/D$. 
Using the data collected in column 6 of the table~\ref{tab:loops:Observables}
we have performed a fit according to those scaling laws. Results are
summarized in table~\ref{tab:loops:FitCvMax}.
\begin{table}[ht]
  \begin{center}
    \begin{tabular}[c]{rr@{.}lrll|lll}
      &  \multicolumn{5}{c|}{1st order, eq.~(\ref{eq:cmax1})} &
      \multicolumn{3}{c}{    2nd order, eq.~(\ref{eq:cmax2})} \\
      \multicolumn{1}{c}{$L$}                 &
      \multicolumn{2}{c}{$a_1$}                &
      \multicolumn{1}{c}{$10^{-4} b_1$}        &
      \multicolumn{1}{c}{$\chi^2/_{d.o.f}$}    &
      \multicolumn{1}{c|}{${\cal Q}$}          &
      \multicolumn{1}{c}{$\alpha/\nu$}         &
      \multicolumn{1}{c}{$\chi^2/_{d.o.f}$}    &
      \multicolumn{1}{c}{${\cal Q}$}           \\
      \hline
      6-24 &  3&16(6) & $2.5(3)$ & $2.9\cdot 10^{2}$  & $0.51\cdot 10^{-375}$
                      & 2.44(4)  & 18.8 & $0.97\cdot 10^{-18}$\\
      8-24 &  5&08(5) & $2.1(3)$ & $7.3\cdot 10$      & $0.38\cdot 10^{-76}$ 
                      & 2.7(7)   & 19.2 & $0.91\cdot 10^{-15}$\\
      10-24 & 6&7(1)  & $1.9(3)$ & 5.1                & $0.43\cdot 10^{-3}$
                      & 3.60(13) & 1.5                & 0.22  \\
      12-24 & 7&7(3)  & $1.8(3)$ & 2.8                & $0.38\cdot 10^{-1}$ 
                      & 3.4(2)   & 1.6                & 0.19  \\
      14-24 & 8&6(4)  & $1.7(5)$ & $5.8\cdot 10^{-2}$ & 0.94 
                      & 3.9(3)   & 0.079              & 0.78  \\
      16-24 & 8&8(9)  & $1.7(6)$ & $6.2\cdot 10^{-3}$ & 0.94 
                      & -        & -                  & -     \\
    \end{tabular}%
    \caption{Fits of $C_v^{max}$. The 1st order fit was done
      using eq.~(\ref{eq:cmax1}) (growing proportional to $L^D$). 2nd order fit
      corresponds to eq.~(\ref{eq:cmax2}), that is, growing as $L^{\alpha/\nu}$,
      being $\alpha/\nu$ free. Notice that in both cases the best fits are
      obtained using only the values $14<L<24$. In addition, the best 2nd
      order fit gives and exponent close to 4.}
    \label{tab:loops:FitCvMax}%
  \end{center}
\end{table}

\figlooopsCvMax

Fig.~\ref{fig:loops:CvMax} shows the measured values of $C_v^{max}$  as a
function of $L^4$. Notice the alignment of the points, exhibiting a 1st order
behavior. Solid line is the 1st order fit using only values $14<L<24$.

\subsection{Critical coupling\label{subsec:loops:critic:BetaCrit}}

The critical coupling must reach their asymptotic behavior obeying the
following laws,
\begin{equation}
\beta_c(L)=\beta_c(\infty)+ A L^{-1/\nu}  \quad \text{(2nd order),}\quad
\beta_c(L)=\beta_c(\infty)+ A L^{-D}      \quad \text{(1st order).}
\label{eq:beta1}
\end{equation}
It is common to say that 1st order transition have a ``critical exponent'' of
$\nu=1/D$ (=0.25 in 4 dimensions). Trivial 2nd order transitions have
$\nu=0.5$. 
Nevertheless, some weak 1st order transitions have an intermediate exponent
and is not easy to elucidate the order of the transition from this scaling
law. In our case, the errors are so large that all the fits are equally
``good'' and therefore this quantity does not provide much 
evidence on the order of the phase transition.

\subsection{Lee-Yang zeroes\label{subsec:loops:critic:LeeYang}}

The zeroes of the partition function analytically extended to the complex plane
manifest a scaling behavior governed by the critical exponent  
$-1/\nu$\cite{Marinari:1996dh,Bowick:1993hn,Marinari:1984wp,Falcioni:1982cz} 
Table~\ref{tab:loops:Observables} show the location of the zeroes closest to
the real axis. From this data, a fit to the imaginary part of  $\beta$ gives
\begin{equation}
  \chi^{(0)}(L) = B L^{-\frac{1}{\nu}}
\end{equation}

This quantity is rather insensitive to the finite size effects. The best fit
($5<L<20$) gives
\begin{equation}
  B   = 2.3(2) \text{\hspace{2cm}}  \nu = 0.329(3)
\end{equation}
with $\chi^2/_{d.o.f}=0.6$ and ${\cal Q}=0.7$. 
Fig~\ref{fig:loops:FitNu} show the data points and the regression curve.

\subsection{Latent heat\label{subsec:loops:critic:DeltaE}}

The last  column of table~\ref{tab:loops:Observables}
show the latent heat for each lattice size.
We expect a behavior as 
\begin{equation}
  \Delta e(L) = \Delta e_1(\infty) + c_1 L^{-d_1}
  \label{eq:loops:gap_power}
\end{equation}
This gives the result  $\Delta e_1(\infty)=0.0117(13)$ with $d_1=1.8(4)$.
($\chi^2/{d.o.f}= 0.8$ i ${\cal Q}=0.53$)
We have also performed a fit as
\begin{equation}
  \Delta e(L) = \Delta e_2(\infty) + c_2 e^{-d_2 L}
  \label{eq:loops:gap_exp}
\end{equation}
with the result $\Delta e_2(\infty)=0.0135(6)$ i $d_2=0.19(3)$ 
($\chi^2/{d.o.f}= 0.5$ i ${\cal Q}=0.75$)

Figure~\ref{fig:loops:gap} show these two different fits together with the
measured data. In both cases and asymptotic behavior with 
 $\Delta e=0$ is ruled out.

\figlooopsFitNuGap

\section{Conclusions\label{sec:loops:conclu}}

Among different appealing features the world-sheet formulation avoids the 
summation 
over redundant gauge configurations. This fact implies an amelioration
of the convergence to the equilibrium. As a consequence, the time series
generated using this approach show a high degree of ``tunneling'' between 
phases, a phenomenon that is depressed in the standard Wilson
formulation.

We have observed in the simulations a standard first order behavior in
several observables (autocorrelations, specific heat, latent heat) 
with an intermediate behavior between the first order
($\nu=0.25$) 
and second order  Gaussian exponent ($\nu=0.5$) like other weak first order
transitions. 
Wether or not this observed first order signature is an artifact of the 
topology/boundary-conditions of the lattice is an issue we will
analyze in a future work.  


\begin{thebibliography}{10}

\bibitem{Campos:1998jp}
I.~Campos, A.~Cruz, and A.~Tarancon, ``A study of the phase transition in 4-d
  pure compact u(1) lgt on toroidal and spherical lattices,''
  \href{http://xxx.lanl.gov/abs/hep-lat/9803007}{{\tt hep-lat/9803007}}.

\bibitem{Campos:1997br}
I.~Campos, A.~Cruz, and A.~Tarancon, ``First order signatures in 4-d pure
  compact u(1) gauge theory with toroidal and spherical topologies,''
  \href{http://xxx.lanl.gov/abs/hep-lat/9711045}{{\tt hep-lat/9711045}}.

\bibitem{Klaus:1997be}
B.~Klaus and C.~Roiesnel, ``High statistics finite size scaling analysis of
  u(1) lattice gauge theory with wilson action,''
  \href{http://xxx.lanl.gov/abs/hep-lat/9801036}{{\tt hep-lat/9801036}}.

\bibitem{Roiesnel:1997am}
C.~Roiesnel, ``Finite size effects at phase transition in compact u(1) gauge
  theory,'' {\em Nucl. Phys. Proc. Suppl.} {\bf 63} (1998) 697,
  \href{http://xxx.lanl.gov/abs/hep-lat/9709081}{{\tt hep-lat/9709081}}.

\bibitem{Jersak:1996mj}
J.~Jersak, C.~B. Lang, and T.~Neuhaus, ``Four-dimensional pure compact u(1)
  gauge theory on a spherical lattice,'' {\em Phys. Rev.} {\bf D54} (1996)
  6909--6922, \href{http://xxx.lanl.gov/abs/hep-lat/9606013}{{\tt
  hep-lat/9606013}}.

\bibitem{Baig:1994ia}
M.~Baig and H.~Fort, ``Fixed boundary conditions and phase transitions in pure
  gauge compact qed,'' {\em Phys. Lett.} {\bf B332} (1994) 428--432,
  \href{http://xxx.lanl.gov/abs/hep-lat/9406003}{{\tt hep-lat/9406003}}.

\bibitem{Banks:1977cc}
T.~Banks, R.~Myerson, and J.~Kogut, ``Phase transitions in abelian lattice
  gauge theories,'' {\em Nucl. Phys.} {\bf B129} (1977) 493.

\bibitem{Baig:1994ib}
M.~Baig, H.~Fort, and J.~B. Kogut, ``Monopole percolation in pure gauge compact
  qed,'' {\em Phys. Rev.} {\bf D50} (1994) 5920--5923,
  \href{http://xxx.lanl.gov/abs/hep-lat/9406004}{{\tt hep-lat/9406004}}.


\bibitem{Lautrup:1980xr}
B.~Lautrup and M.~Nauenberg, ``Phase transition in four-dimensional compact
  qed,'' {\em Phys. Lett.} {\bf 95B} (1980) 63.

\bibitem{Bhanot:1981zg}
G.~Bhanot, ``The nature of the phase transition in compact qed,'' {\em Phys.
  Rev.} {\bf D24} (1981) 461.

\bibitem{Jersak:1983yz}
J.~Jersak, T.~Neuhaus, and P.~M. Zerwas, ``U(1) lattice gauge theory near the
  phase transition,'' {\em Phys. Lett.} {\bf 133B} (1983) 103.

\bibitem{Azcoiti:1990ue}
V.~Azcoiti, G.~di~Carlo, and A.~F. Grillo, ``Towards a precise determination of
  the order of the phase transition in compact pure gauge qed,'' {\em Phys.
  Lett.} {\bf B238} (1990) 355.

\bibitem{Bhanot:1992nf}
G.~Bhanot, T.~Lippert, K.~Schilling, and P.~Uberholz, ``First order transitions
  and the multihistogram method,'' {\em Nucl. Phys.} {\bf B378} (1992)
  633--651.

\bibitem{Gambini:1980wm}
R.~Gambini and A.~Trias, ``Second quantization of the free electromagnetic
  field as quantum mechanics in the loop space,'' {\em Phys. Rev.} {\bf D22}
  (1980) 1380.

\bibitem{Gambini:1986ew}
R.~Gambini and A.~Trias, ``Gauge dynamics in the c representation,'' {\em Nucl.
  Phys.} {\bf B278} (1986) 436.

\bibitem{Aroca:1994yb}
J.~M. Aroca, M.~Baig, and H.~Fort, ``The lagrangian loop representation of
  lattice u(1) gauge theory,'' {\em Phys. Lett.} {\bf B336} (1994) 54--61,
  \href{http://xxx.lanl.gov/abs/hep-th/9407170}{{\tt hep-th/9407170}}.

\bibitem{Aroca:1996cs}
J.~M. Aroca, M.~Baig, H.~Fort, and R.~Siri, ``Matter fields in the lagrangian
  loop representation: Scalar qed,'' {\em Phys. Lett.} {\bf B366} (1996)
  416--420, \href{http://xxx.lanl.gov/abs/hep-th/9507124}{{\tt
  hep-th/9507124}}.

\bibitem{Polyakov:1979gp}
A.~M. Polyakov, ``String representations and hidden symmetries for gauge
  fields,'' {\em Phys. Lett.} {\bf 82B} (1979) 247.

\bibitem{Polyakov:1980ca}
A.~M. Polyakov, ``Gauge fields as rings of glue,'' {\em Nucl. Phys.} {\bf B164}
  (1980) 171.

\bibitem{Makeenko:1979pb}
Y.~M. Makeenko and A.~A. Migdal, ``Exact equation for the loop average in
  multicolor qcd,'' {\em Phys. Lett.} {\bf 88B} (1979) 135.

\bibitem{Yang:1974kj}
C.~N. Yang, ``Integral formalism for gauge fields,'' {\em Phys. Rev. Lett.}
  {\bf 33} (1974) 445.

\bibitem{Guth:1980gz}
A.~H. Guth, ``Existence proof of a nonconfining phase in four-dimensional u(1)
  lattice gauge theory,'' {\em Phys. Rev.} {\bf D21} (1980) 2291.

\bibitem{Weingarten:1980gn}
D.~Weingarten, ``A lattice field theory for interacting strings,'' {\em Phys.
  Lett.} {\bf 90B} (1980) 280.

\bibitem{Sokal:1989ea}
A.~D. Sokal, ``Monte carlo mehtods in statistical mechanics: Foundations and
  new algorithms,''.

\bibitem{Marinari:1996dh}
E.~Marinari, ``Optimized monte carlo methods,''
  \href{http://xxx.lanl.gov/abs/cond-mat/9612010}{{\tt cond-mat/9612010}}.

\bibitem{Bowick:1993hn}
M.~Bowick, P.~Coddington, L.~ping Han, G.~Harris, and E.~Marinari, ``The phase
  diagram of fluid random surfaces with extrinsic curvature,'' {\em Nucl.
  Phys.} {\bf B394} (1993) 791--821,
  \href{http://xxx.lanl.gov/abs/hep-lat/9209020}{{\tt hep-lat/9209020}}.

\bibitem{Marinari:1984wp}
E.~Marinari, ``Complex zeros of the d = 3 ising model: Finite size scaling and
  critical amplitudes,'' {\em Nucl. Phys.} {\bf B235} (1984) 123.

\bibitem{Falcioni:1982cz}
M.~Falcioni, E.~Marinari, M.~L. Paciello, G.~Parisi, and B.~Taglienti,
  ``Complex zeros in the partition function of the four- dimensional su(2)
  lattice gauge model,'' {\em Phys. Lett.} {\bf 108B} (1982) 331.

\end{thebibliography}

\vspace{5mm}


\section*{Acknowledgements}
This work has been partially supported by
research project CICYT AEN95/0882.
H.F. was supported in part by CONICYT Project Nr 318 and CSIC. 

\providecommand{\href}[2]{#2}\begingroup\raggedright\endgroup

\end{document}